\documentclass[twocolumn,showpacs,preprintnumbers,amsmath,amssymb,prb]{revtex4}

\usepackage{graphicx}
\usepackage{dcolumn}
\usepackage{bm}
\usepackage{color}
\newcommand{\ua}{\uparrow}
\newcommand{\da}{\downarrow}
\newcommand{\dg}{\dagger}

\begin{document}

\preprint{}

\title{Theoretical study of insulating mechanism in multi-orbital Hubbard models with a large spin-orbit coupling: Slater versus Mott scenario in Sr$_2$IrO$_4$}

\author{Hiroshi Watanabe$^{1,2}$}
 \email{h-watanabe@riken.jp}
\author{Tomonori Shirakawa$^{2,3,4}$}
\author{Seiji Yunoki$^{1,2,3,4}$}
\affiliation{%
$^1$Computational Quantum Matter Reasearch Team, RIKEN Center for Emergent Matter Science (CEMS), Wako, Saitama 351-0198, Japan\\
$^2$CREST, Japan Science and Technology Agency, Kawaguchi, Saitama 332-0012, Japan\\
$^3$Computational Condensed Matter Physics Laboratory, RIKEN, Wako, Saitama 351-0198, Japan\\
$^4$Computational Materials Science Research Team, RIKEN Advanced Institute for Computational Science (AICS), Kobe, Hyogo 650-0047, Japan
}%

\date{\today}

\begin{abstract}
To examine the insulating mechanism of 5$d$ transition metal oxide Sr$_2$IrO$_4$, we study the ground state properties of 
a three-orbital Hubbard model with a large relativistic spin-orbit coupling on a square lattice. 
Using a variational Monte Carlo method, 
we find that the insulating state appearing in the ground state phase diagram for one hole per site 
varies from a weakly correlated to a strongly correlated antiferromagnetic (AF) state with increasing Coulomb interactions. 
This crossover is characterized by the different energy gain mechanisms of the AF insulating state, i.e., from 
an interaction-energy driven Slater-type insulator to a band-energy driven Mott-type insulator with 
increasing Coulomb interactions. 
Our calculations reveal that Sr$_2$IrO$_4$ is a ``moderately correlated" AF insulator located in the intermediate coupling region 
between a Slater-type and a Mott-type insulators.
\end{abstract}

\maketitle

\section{Introduction}
The 3$d$ transition metal oxides have been extensively studied as typical examples of strongly correlated electron systems.
One of the most fascinating features is represented by a Mott insulator which emerges as a result of strong electron correlations 
beyond the single-particle band theory.~\cite{Imada,Phillips} 
The most extensively studied systems are high-temperature cuprate superconductors 
where high-temperature superconductivity (SC) has been observed by introducing mobile carriers into Mott insulators~\cite{Bednorz} 
and subsequently various novel concepts have been proposed including 
unconventional SC,~\cite{Scalapino,Moriya,Balatsky,Eschrig,Edegger} 
various symmetry broken orders,~\cite{Kivelson,Demler,Sachdev} and pseudogap 
phenomena.~\cite{Timusk,Sadovskii,Norman,PLee,Tremblay,Heufner,Rice} 
The study of strongly correlated electron systems enlarges our fundamental knowledge of quantum states of matters and  
therefore search for novel Mott insulators is valuable for further progress in condensed matter physics. 

Recently, 5$d$ transition metal oxide Sr$_2$IrO$_4$ with the layered perovskite structure~\cite{Randall,Cao} has 
attracted much attention as a candidate for a novel 
Mott insulator.~\cite{Kim1,Kim2,Jackeli,Jin,Watanabe1} 
In this material, three $t_{2g}$ orbitals of Ir atom are hybridized among themselves with a large relativistic spin-orbit coupling (SOC), inherent in $5d$ transition metal,  
and are occupied nominally by five $5d$ electrons. Because of this quantum entanglement of spin and orbital degrees of freedom, 
an effective total angular momentum $J_{\mathrm{eff}}$=$|-\bm{L}+\bm{S}|=1/2$ state is stabilized locally 
at each Ir atom.~\cite{Sugano} The Coulomb interactions are generally believed to be smaller compared with 
the whole band width of the bands formed by $t_{2g}$ orbitals in 5$d$ systems. However, 
when the $J_{\mathrm{eff}}=1/2$ state forms a band and this band is detached from the rest of the bands due to a large SOC, 
the width of the $J_{\mathrm{eff}}=1/2$ band is narrower and 
comparable to the Coulomb interactions, which thus can induce a novel $J_{\mathrm{eff}}=1/2$ insulating state. 
This picture is indeed supported in Sr$_2$IrO$_4$ both experimentally~\cite{Kim1,Kim2,Ishii,Fujiyama1} 
and theoretically.~\cite{Jackeli,Jin,Watanabe1,Shirakawa,Onishi} 
Further experiments have revealed various interesting properties of this insulator~\cite{Kim3,Lee,Cetin,Qi,Dessau,Clancy,Fujiyama2}
and its Ba counterpart of Ba$_2$IrO$_4$.~\cite{Okabe1,Boseggia,Okabe2,Moser,Sala}
Moreover, theoretical studies have proposed several novel features such as possible SC and 
a topological insulator.~\cite{Wang,Kim4,Carter,Katukuri,Watanabe2,Yang} 
However, the origin of the $J_{\mathrm{eff}}=1/2$ insulating state is still under debate. While the Mott-type 
mechanism (i.e., a strongly correlated insulator) has been originally proposed,~\cite{Kim1,Kim2,Ishii} recent reports have 
suggested the Slater-type mechanism (i.e., a weakly correlated insulator)~\cite{Kini,Arita,Li,Suga} as well as both characters 
coexisting in Sr$_2$IrO$_4$.~\cite{Hsieh}
The difficulty of this problem is due to the fact that the Coulomb interactions and the antiferromagnetic (AF) order can both 
split the band and make the system insulating. The aim of this paper is to clarify the insulating mechanism of Sr$_2$IrO$_4$ 
from the microscopic point of view by considering the energy gain mechanism.

In a Mott insulator, the Coulomb interactions are responsible for the insulating behavior and the insulating gap is determined 
essentially by the energy difference between the upper and lower Hubbard bands, where the AF
order has only a secondary effect. 
To the contrary, in a Slater insulator, the translational symmetry breaking AF order induces the insulating behavior and 
thus very often the band structure is essential for its stability. 
In the experiments on Sr$_2$IrO$_4$, the temperature dependence of the resistivity is found insulating up to 600~K and 
no significant change is observed at the N{\'e}el temperature,~\cite{Chikara} strongly suggesting that Sr$_2$IrO$_4$ is 
a Mott-type insulator. Theoretically, however, the insulating mechanism in a multi-orbital Hubbard system with a large SOC has 
not been explored in general and therefore the systematic study is highly desired.

In this paper, using a variational Monte Carlo (VMC) method, we study the ground state properties of a three-orbital Hubbard model 
with a large SOC on the square lattice 
and examine the insulating mechanism of Sr$_2$IrO$_4$.
We find that in the ground state phase diagram with one hole per site the insulating state is always AF ordered and 
changes its character from a weakly correlated (Slater-type) insulator to a 
strongly correlated (Mott-type) insulator as the Coulomb interactions increase. 
These insulating states are differentiated in the energy gain mechanism which favors the AF insulator over a paramagnetic state, i.e., 
an interaction-energy driven Slater-type insulator and a band-energy driven Mott-type insulator. 
We also find that there exists an intermediate region where both energy gain mechanisms work due to the strong renormalization 
of the paramagnetic metallic state. Based on our results, we attribute Sr$_2$IrO$_4$ to be 
a ``moderately correlated" AF insulator located between a Slater-type and a Mott-type insulators.

The rest of this paper is organized as follows. In Sec.~\ref{model}, a three-orbital Hubbard model on the two-dimensional square 
lattice is introduced as a low energy effective model for Sr$_2$IrO$_4$. The detailed explanation of the VMC method and 
the variational wave functions are also given in Sec.~\ref{model}. 
The numerical results are then provided in Sec.~\ref{results}. 
In Sec.~\ref{para}, the results for the paramagnetic state are shown and the 
metal-insulator transition within the paramagnetic state is discussed. Calculating 
several physical quantities, we show that the metallic state is strongly renormalized near the metal-insulator transition. 
In Sec.~\ref{af}, the AF state is considered and the energy gain mechanism stabilizing the AF insulator over the paramagnetic 
state is investigated. Systematically analyzing each term in the Hamiltonian, we show that the main sources of the energy gain 
which favors the AF insulator are the interaction terms for small Coulomb interactions 
and the kinetic terms for larger Coulomb 
interactions. We assign Sr$_2$IrO$_4$ to be a ``moderately correlated" AF insulator located in the intermediate coupling region. 
Finally, Sec.~\ref{summary} summaries this paper. 

\section{Model and method}\label{model}

\subsection{Three-orbital Hubbard model}

We consider a three-orbital Hubbard model on the two-dimensional square lattice defined by the following Hamiltonian 
\begin{equation}
H=H_{\mathrm{kin}}+H_{\mathrm{SO}}+H_{\mathrm{I}}, 
\label{mh}
\end{equation}
where the kinetic term $H_{\mathrm{kin}}$ is described by 
\begin{equation}
H_{\mathrm{kin}}=\sum_{\bm{k},\alpha,\sigma}\varepsilon_{\alpha}(\bm{k})c_{\bm{k}\alpha\sigma}^{\dg}
        c_{\bm{k}\alpha\sigma}, 
\end{equation}
the SOC term $H_{\mathrm{SO}}$ with a coupling constant $\lambda$ is given as 
\begin{equation}
H_{\mathrm{SO}}=\lambda\sum_{i}\sum_{\alpha,\beta}\sum_{\sigma,\sigma'}\langle\alpha|\bm{L}_i|\beta\rangle\cdot\langle\sigma|\bm{S}_i|\sigma'\rangle c^\dag_{i\alpha\sigma}c_{i\beta\sigma'}, 
\end{equation}
and the Coulomb interaction terms $H_{\mathrm{I}}$ are composed of four terms, 
\begin{equation}
H_{\mathrm{I}} = H_U + H_{U'} + H_J + H_{J'}, 
\label{int}
\end{equation}
i.e., the intra-orbital interaction term, 
\begin{equation}
H_U = U\sum_{i,\alpha}n_{i\alpha\ua}n_{i\alpha\da}, 
\end{equation}
the inter-orbital interaction term, 
\begin{equation}
H_{U'} =U' \sum_{i,\alpha<\beta,\sigma}\left(n_{i\alpha\sigma}n_{i\beta\bar{\sigma}}
  +n_{i\alpha\sigma}n_{i\beta\sigma}\right), \label{U'}
\end{equation}
the Hund's coupling term,  
\begin{equation}
H_J = J\sum_{i,\alpha<\beta}\left [ (c^{\dg}_{i\alpha\ua}c^{\dg}_{i\beta\da}
  c_{i\alpha\da}c_{i\beta\ua} +\mathrm{H.c.}) 
  - \sum_{\sigma} n_{i\alpha\sigma}n_{i\beta\sigma}\right],  
\end{equation}
and the pair-hopping term, 
\begin{equation}
H_{J'} = J'\sum_{i,\alpha<\beta}(c^{\dg}_{i\alpha\ua}c^{\dg}_{i\alpha\da}
   c_{i\beta\da}c_{i\beta\ua}+\mathrm{H.c.}). 
\end{equation}
Here, $c_{i\alpha\sigma}^\dag$ is a creation operator of electron at site $i$ with spin $\sigma\,(=\ua,\da)$ and orbital 
$\alpha\, (=yz,zx,xy)$ corresponding to three $t_{2g}$ orbitals ($d_{yz}$, $d_{zx}$, and $d_{xy}$), 
and ${\bm L}_i$ (${\bm S}_i$) is orbital (spin) angular momentum operator at site $i$.
The opposite spin of $\sigma$ is indicated by $\bar\sigma$ and $n_{i\alpha\sigma}=c^\dag_{i\alpha\sigma}c_{i\alpha\sigma}$. 
The Fourier transform of $c_{i\alpha\sigma}^\dag$ is given as
\begin{equation}
c_{\bm{k}\alpha\sigma}^\dag = {\frac{1}{\sqrt{N}}}\sum_i e^{i{\bm k}\cdot{\bm r}_i} c_{i\alpha\sigma}^\dag, 
\end{equation}
where $N$ is the total number of sites and ${\bm r}_i$ is the position vector of site $i$.  
In the following, we set $J'=J$ and $U=U'+2J$,~\cite{Kanamori} unless otherwise stated. 

The kinetic and the SOC terms can be combined, $H_0(t_i,\mu_{xy}, \lambda)=H_{\mathrm{kin}}+H_{\mathrm{SO}}$, in the matrix form 
\begin{align}\label{h0}
H_0&=\sum_{\bm{k},\sigma}\left(c^{\dg}_{\bm{k}yz\sigma},c^{\dg}_{\bm{k}zx\sigma},c^{\dg}_{\bm{k}xy\bar{\sigma}}\right) \notag \\
&\;\;\;\;\;\times
\begin{pmatrix}
\varepsilon_{yz}(\bm{k}) & \mathrm{i}s_\sigma\lambda/2 & -s_\sigma\lambda/2 \\
-\mathrm{i}s_\sigma\lambda/2 & \varepsilon_{zx}(\bm{k}) & \mathrm{i}\lambda/2 \\
-s_\sigma\lambda/2 & -\mathrm{i}\lambda/2 & \varepsilon_{xy}(\bm{k})
\end{pmatrix}
\begin{pmatrix}
c_{\bm{k}yz\sigma} \\
c_{\bm{k}zx\sigma} \\
c_{\bm{k}xy\bar{\sigma}}
\end{pmatrix} \\ \notag
&=\sum_{\bm{k},m, s}E_m(\bm{k})a^{\dg}_{\bm{k}m s}
a_{\bm{k}m s},  
\end{align}
where $s_\sigma=1\,(-1)$ for $\sigma=\ua(\da)$.  
Notice that the SOC mixes the different electron spins ($\sigma$ and $\bar{\sigma}$), and the new quasiparticles, 
obtained by diagonalizing $H_0$, are characterized by band index $m\,(=1,2,3)$ 
and pseudospin $s\,(=\ua,\da)$ with a creation operator $a^{\dg}_{\bm{k}m s}$. 
In the atomic limit with $\varepsilon_{yz}(\bm{k})=\varepsilon_{zx}(\bm{k})=\varepsilon_{xy}(\bm{k})=0$, 
the sixfold degenerate $t_{2g}$ levels are split into twofold degenerate $J_{\mathrm{eff}}=1/2$ states ($m=1$) 
and fourfold degenerate $J_{\mathrm{eff}}=3/2$ states ($m=2,3$).~\cite{Sugano} Since the $J_{\mathrm{eff}}=1/2$ states 
are higher in energy than the $J_{\mathrm{eff}}=3/2$ states, 
all states but the $J_{\mathrm{eff}}=1/2$ states are fully occupied for electron density $n=5$, i.e., one hole per site. 

\begin{figure}[t]
\begin{center}
\includegraphics[width=0.9\hsize]{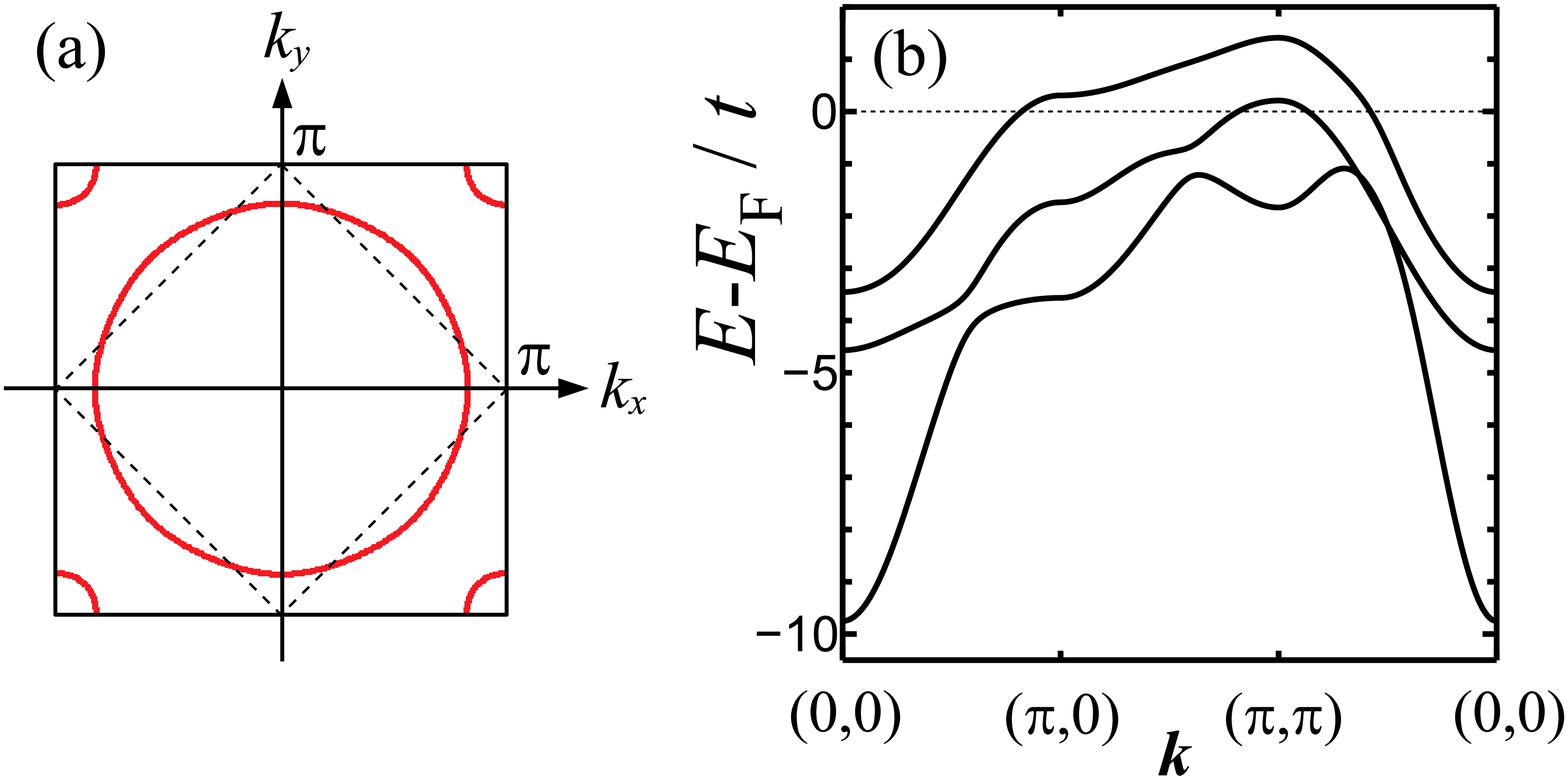}
\caption{\label{fig1} 
(color online) 
(a) Fermi surfaces and (b) energy dispersions of the non-interacting tight-binding energy band 
with electron density $n=5$. The SOC is set to be $\lambda=1.028t$ ($\sim 0.37$ eV). The other parameters 
are given in Eq.~(\ref{tb_parameter}). 
Dashed lines in (a) represent the folded AF Brillouin zone. $E_{\rm F}$ stands for Fermi energy. 
} 
\end{center}
\end{figure}

\begin{figure}[t]
\begin{center}
\includegraphics[width=0.9\hsize]{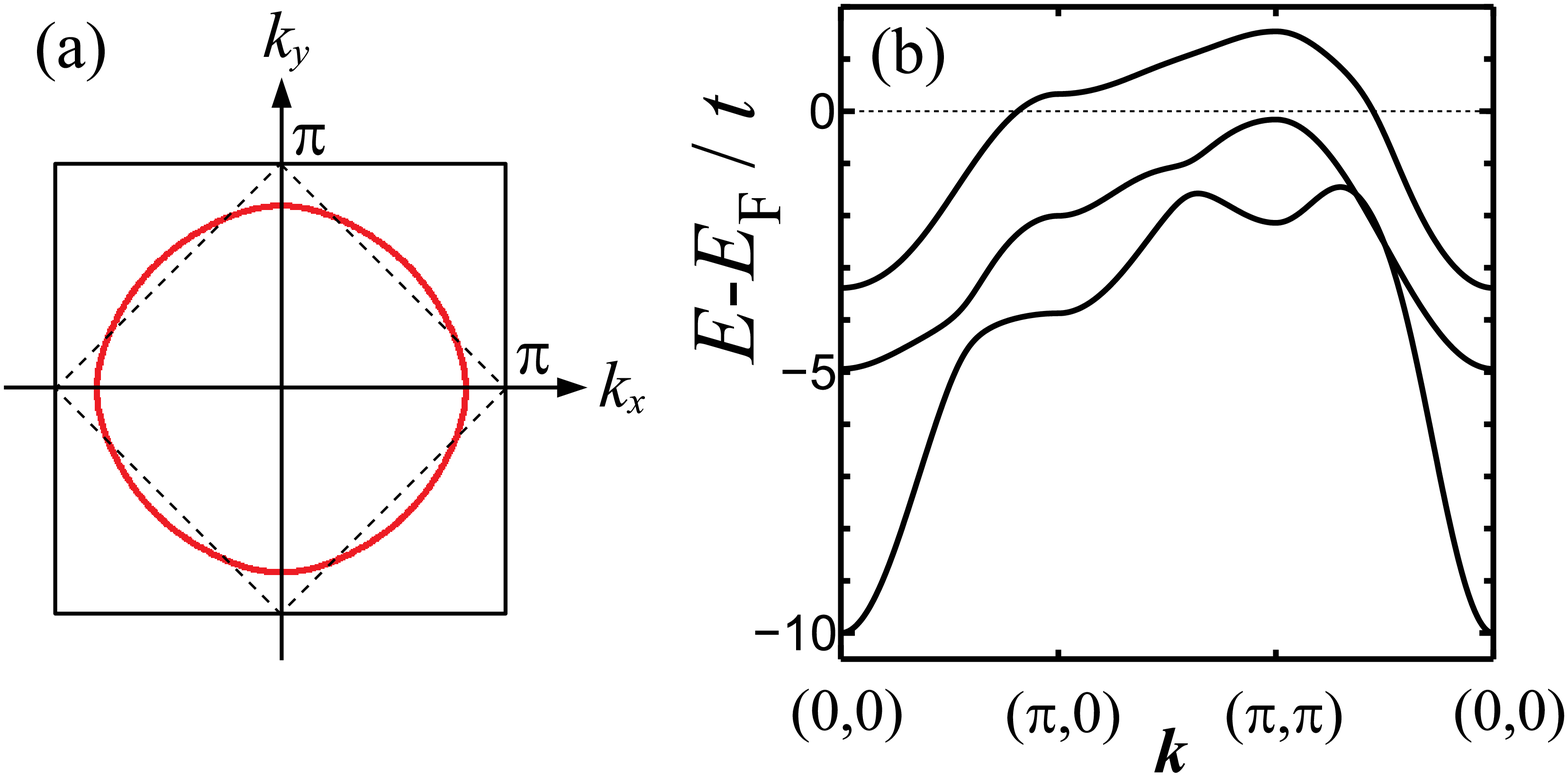}
\caption{\label{fig2}
(color online) 
Same as Fig.~\ref{fig1} but the SOC is set to be $\lambda=1.4t$ ($\sim0.50$ eV).  
} 
\end{center}
\end{figure}

In Refs.~\onlinecite{Watanabe1} and \onlinecite{Watanabe2}, 
we have constructed the non-interacting tight-binding energy band for Sr$_2$IrO$_4$:
\begin{align}
\varepsilon_{yz}(\bm{k})=&-2t_5\cos k_x-2t_4\cos k_y, \\
\varepsilon_{zx}(\bm{k})=&-2t_4\cos k_x-2t_5\cos k_y, \\
\varepsilon_{xy}(\bm{k})=&-2t_1(\cos k_x+\cos k_y)-4t_2\cos k_x\cos k_y \notag \\
                                           &-2t_3(\cos 2k_x+\cos 2k_y)+\mu_{xy}
\end{align}
with a set of tight-binding parameters 
\begin{eqnarray}\label{tb_parameter}
&&(t_1, t_2, t_3, t_4, t_5, \mu_{xy})\nonumber \\
&&\quad\quad\quad=(0.36, 0.18, 0.09, 0.37, 0.06, -0.36)\,{\rm eV}.
\end{eqnarray} 
In the following, $t_1\equiv t$ is used for an energy unit.
To study the effect of the SOC, we choose two different values of 
$\lambda$, i.e., $\lambda=1.028t$ ($\sim0.37$ eV) and $1.4t$ ($\sim0.50$ eV), both of which are 
within the range of realistic values for Sr$_2$IrO$_4$.
The corresponding Fermi surface (FS) and energy dispersions are shown in Fig.~\ref{fig1} and Fig.~\ref{fig2}. 
The noticeable features are summarized as follows: 
(i) the topmost band, i.e., $J_{\rm eff}=1/2$ band with $m=1$, is detached from the other two bands, i.e., 
$J_{\rm eff}=3/2$ bands with $m=2$ and $3$, 
(ii) the separation in energy between the topmost band and the other two bands increases with $\lambda$, and 
(iii) hole pockets appearing at ${\bm k}=(\pi,\pi)$ (and the equivalent momenta) for $\lambda=1.028t$ disappears 
as $\lambda$ increases and a single circular-like FS is formed by the topmost band for larger $\lambda$. 

\subsection{Variational Monte Carlo (VMC) method}\label{vmc}

The effect of Coulomb interactions is treated using a VMC method.~\cite{Watanabe1,Watanabe2} 
The trial wave function $\left|\Psi \right>$ considered here is composed of three parts,  
\begin{equation}
\left|\Psi \right>=P_{\mathrm{J_c}}P^{(3)}_{\mathrm{G}}\left|\Phi \right>.
\label{wf}
\end{equation}
The one-body part $\left|\Phi \right>$ is described by the ground state of 
$\tilde H_0= H_0(\tilde{t}_i, \tilde{\mu}_{xy}, \tilde{\lambda}_{\alpha\beta})$ 
with variational ``renormalized" tight-binding parameters $\{ \tilde{t}_i, \tilde{\mu}_{xy}, \tilde{\lambda}_{\alpha\beta} \}$, 
where $H_0$ is given in Eq.~(\ref{h0}). 
Notice that we introduce an orbital dependent ``effective'' SOC constant: $\lambda\rightarrow\tilde{\lambda}_{\alpha\beta}$.

To treat magnetically ordered states, a term with a different magnetic order parameter is added to $\tilde H_0$.
Here, we consider two different magnetic orders, i.e., out-of-plane AF order (along $z$ axis, $z$-AF) described by 
\begin{equation}
\tilde H_{\rm AF}^z(\tilde M^z_1,\tilde M^z_2,\tilde M^z_3)=\sum_{i,m} \tilde M^z_m\mathrm{e}^{\mathrm{i}\bm{Q}\cdot\bm{r}_i}(a^{\dg}_{im\ua}a_{im\ua}
-a^{\dg}_{im\da}a_{im\da})
\end{equation}
and in-plane AF order (along $x$ axis, $x$-AF) described by
\begin{equation}
\tilde H_{\rm AF}^x(\tilde M^x_1,\tilde M^x_2,\tilde M^x_3)=\sum_{i,m} \tilde M^x_m\mathrm{e}^{\mathrm{i}\bm{Q}\cdot\bm{r}_i}(a^{\dg}_{im\ua}a_{im\da}
+a^{\dg}_{im\da}a_{im\ua}), 
\end{equation} 
where $a^{\dg}_{ims}$ is the Fourier transformation of $a^{\dg}_{\bm{k}ms}$ and $\bm{Q}=(\pi,\pi)$. 
The order parameters ($\tilde M^z_1,\tilde M^z_2,\tilde M^z_3$) for $z$-AF and ($\tilde M^x_1,\tilde M^x_2,\tilde M^x_3$) 
for $x$-AF are variational parameters to be optimized. 
With an appropriate basis transformation, we obtain the original $t_{2g}$ orbital representation in real space 
and construct the Slater determinant $\left|\Phi \right>$ with 5 electrons (i.e., 1 hole) per site for VMC simulations.

\begin{table*}
\caption{The Gutzwiller parameters $g_\gamma$ in the Gutzwiller operator [Eq.~(\ref{go})] 
extended for the three-orbital system. 64 different local electron configurations 
$|\gamma\rangle\, (\gamma=0, 1, \cdots, 63)$ are composed of 4 different states, $0$, $\ua$, $\da$, and $\ua\da$, 
for each orbital $d_{yz}$, $d_{zx}$, and $d_{xy}$ indicated in the second column. The corresponding Hartree 
energy $E_{\rm I}=\left<\gamma|H_{\mathrm{I}}|\gamma\right>$ is given in the third column. 
These 64 local electron configurations are divided into 12 groups such that the electron configurations in the same group 
have the same $E_{\rm I}$. We assume that the Gutzwiller parameters $g_\gamma$ in the same group 
have the same value, indicated in the fourth column by $w_i$. 
Note that $\gamma$ in the first column is given by representing the local electron configurations in quaternary notation, i.e., 
$\gamma=m_{d_{yz}}4^2+m_{d_{zx}}4^1+m_{d_{xy}}4^0$ where $m_{d_\alpha}=0,1,2$, and $3$ for electron configurations of 
orbital $\alpha$ with $0$, $\ua$, $\da$, and $\ua\da$, respectively. 
} \label{Gutzwiller}

\begin{tabular}{rr}
\begin{minipage}{0.35\hsize}
\[
\begin{array}{c|ccc|c|c}
\hline
\gamma & d_{yz} & d_{zx} & d_{xy} & E_{\rm I} & g_{\gamma} \\ \hline\hline
0 & 0 & 0 & 0 & 0 & w_1\\ 
1 & 0 & 0 & \ua & 0\\
2 & 0 & 0 & \da & 0\\
4 & 0 & \ua & 0 & 0\\
8 & 0 & \da & 0 & 0 \\
16 & \ua & 0 & 0 & 0 \\
32 & \da & 0 & 0 & 0 \\ \hline
3 & 0 & 0 & \ua\da & U & w_2 \\
12 & 0 & \ua\da & 0 & U \\
48 & \ua\da & 0 & 0 & U \\ \hline
5 & 0 & \ua & \ua & U'-J & w_3\\
10 & 0 & \da & \da & U'-J \\
17 & \ua & 0 & \ua & U'-J \\
20 & \ua & \ua & 0 & U'-J \\
34 & \da & 0 & \da & U'-J \\
40 & \da & \da & 0 & U'-J \\ \hline
6 & 0 & \ua & \da & U' & w_4\\
9 & 0 & \da & \ua & U' \\
18 & \ua & 0 & \da & U' \\
24 & \ua & \da & 0 & U' \\
33 & \da & 0 & \ua & U' \\
36 & \da & \ua & 0 & U' \\ \hline
7 & 0 & \ua & \ua\da & U+2U'-J & w_5 \\
11 & 0 & \da & \ua\da & U+2U'-J \\
13 & 0 & \ua\da & \ua & U+2U'-J \\
14 & 0 & \ua\da & \da & U+2U'-J \\
19 & \ua & 0 & \ua\da & U+2U'-J \\
28 & \ua & \ua\da & 0 & U+2U'-J \\
35 & \da & 0 & \ua\da & U+2U'-J \\
44 & \da & \ua\da & 0 & U+2U'-J \\
49 & \ua\da & 0 & \ua & U+2U'-J \\
50 & \ua\da & 0 & \da & U+2U'-J \\ \hline
\end{array}
\]
\end{minipage}

\begin{minipage}{0.35\hsize}
\[
\begin{array}{c|ccc|c|c}
\hline
\gamma & d_{yz} & d_{zx} & d_{xy} & E_{\rm I} &g_{\gamma} \\ \hline\hline
52 & \ua\da & \ua & 0 & U+2U'-J & w_5\\
56 & \ua\da & \da & 0 & U+2U'-J \\ \hline
21 & \ua & \ua & \ua & 3U'-3J & w_6 \\
42 & \da & \da & \da & 3U'-3J \\ \hline
22 & \ua & \ua & \da & 3U'-J & w_7 \\
25 & \ua & \da & \ua & 3U'-J \\
26 & \ua & \da & \da & 3U'-J \\ 
37 & \da & \ua & \ua & 3U'-J \\
38 & \da & \ua & \da & 3U'-J \\
41 & \da & \da & \ua & 3U'-J \\ \hline
15 & 0 & \ua\da & \ua\da & 2U+4U'-2J & w_8 \\
51 & \ua\da & 0 & \ua\da & 2U+4U'-2J \\
60 & \ua\da & \ua\da & 0 & 2U+4U'-2J \\ \hline
23 & \ua & \ua & \ua\da & U+5U'-3J & w_9 \\
29 & \ua & \ua\da & \ua & U+5U'-3J \\
43 & \da & \da & \ua\da & U+5U'-3J \\
46 & \da & \ua\da & \da & U+5U'-3J \\
53 & \ua\da & \ua & \ua & U+5U'-3J \\
58 & \ua\da & \da & \da & U+5U'-3J \\ \hline 
27 & \ua & \da & \ua\da & U+5U'-2J & w_{10} \\
30 & \ua & \ua\da & \da & U+5U'-2J \\
39 & \da & \ua & \ua\da & U+5U'-2J \\
45 & \da & \ua\da & \ua & U+5U'-2J \\
54 & \ua\da & \ua & \da & U+5U'-2J \\
57 & \ua\da & \da & \ua & U+5U'-2J \\ \hline
31 & \ua & \ua\da & \ua\da & 2U+8U'-4J & w_{11} \\
47 & \da & \ua\da & \ua\da & 2U+8U'-4J \\ 
55 & \ua\da & \ua & \ua\da & 2U+8U'-4J \\
59 & \ua\da & \da & \ua\da & 2U+8U'-4J \\
61 & \ua\da & \ua\da & \ua & 2U+8U'-4J \\
62 & \ua\da & \ua\da & \da & 2U+8U'-4J \\ \hline
63 & \ua\da & \ua\da & \ua\da & 3U+12U'-6J & w_{12} \\ \hline
\end{array}
\]
\end{minipage}
\end{tabular}

\end{table*}

The Gutzwiller operator 
\begin{equation}
P^{(3)}_{\mathrm{G}}=\prod_{i,\gamma}\left[1-(1-g_\gamma)\left|\gamma\right>\left<\gamma\right|_i\right]
\label{go}
\end{equation} 
in $\left|\Psi \right>$ is the one extended for the three-orbital system.~\cite{Watanabe1,Watanabe2} 
Here, $i$ is a site index and $\gamma$ represents possible electron configurations at each site, namely, 
$\left|0\right>=\left|0\;0\;0\right>$, $\left|1\right>=\left|0\;0\ua\right>$, $\cdots$, 
$\left|63\right>=\left|\ua\da\;\ua\da\;\ua\da\right>$.
The variational parameters $g_\gamma$'s vary from 0 
to 1, which control the weight of each electron configuration.  
Here, we classify the possible 64 local electron configurations into 12 groups by the local Coulomb interaction energy 
$E_{\rm I}=\langle\gamma|H_{\rm I}|\gamma\rangle$, 
and set the same value of $g_\gamma$'s for electron configurations with the same $E_{\rm I}$. 
The explicit grouping is shown in TABLE~\ref{Gutzwiller}.

The remaining operator 
\begin{equation}\label{jastrow}
P_{\mathrm{J_c}}=\exp\left[-\sum_{i\neq j}v_{ij}n_in_j\right]
\end{equation}
in $\left|\Psi \right> $ is the charge Jastrow factor, which controls the long-range charge correlations. 
Here, $n_i=\sum_{\alpha\sigma}n_{i\alpha\sigma}$ is the electron number at site $i$. 
We assume that $v_{ij}$ depends only on the distances, $v_{ij}=v(|\bm{r}_i-\bm{r}_j|)$, and consider 
the range of $r=|\bm{r}_i-\bm{r}_j|<L/2$ for a square lattice of $N=L\times L$. 
The numbers of independent variational parameters $v_{ij}$ are, for example, 29 for $L=16$ and 43 for $L=20$.

The ground state energies are calculated with a VMC method.
The variational parameters, as many as 80 parameters for a 20$\times$20 square lattice, are 
simultaneously optimized to minimize the variational energy 
by using the stochastic reconfiguration method.~\cite{Sorella} 
We employ periodic and antiperiodic boundary conditions in $x$ and $y$ directions, respectively.~\cite{Shiba} 
The largest system size treated in this paper is $L=20$.

\section{Results}\label{results}

\subsection{Paramagnetic state and metal-insulator transition}\label{para}

In the previous paper,~\cite{Watanabe1} we have found that the ground state of the three-orbital Hubbard model 
given in Eq.~(\ref{mh}) for Sr$_2$IrO$_4$ is well described by the Gutzwiller-Jastrow type wave function $|\Psi\rangle$ 
[Eq.~(\ref{wf})] with in-plane AF order (see in Sec.~\ref{vmc}). 
Before discussing this AF state in detail, let us first focus on a paramagnetic state and examine the metal-insulator transition 
in the paramagnetic phase. 
Since the electron density per unit cell is an odd integer ($n=5$), this transition should be a ``true" Mott transition 
without breaking translational symmetry by any magnetic order. 
As shown below, the FS deformation due to the renormalization of the one-body parameters 
$\{ \tilde{t}_i, \tilde{\mu}_{xy}, \tilde{\lambda}_{\alpha\beta} \}$ in $|\Phi\rangle$ is quite important for describing the 
metal-insulator transition. Therefore, care must be taken in considering all possible FSs in the wave functions 
by properly occupying $\bm{k}$ points with electrons.~\cite{note1}

\begin{figure}[thbp]
\begin{center}
\includegraphics[width=0.8\hsize]{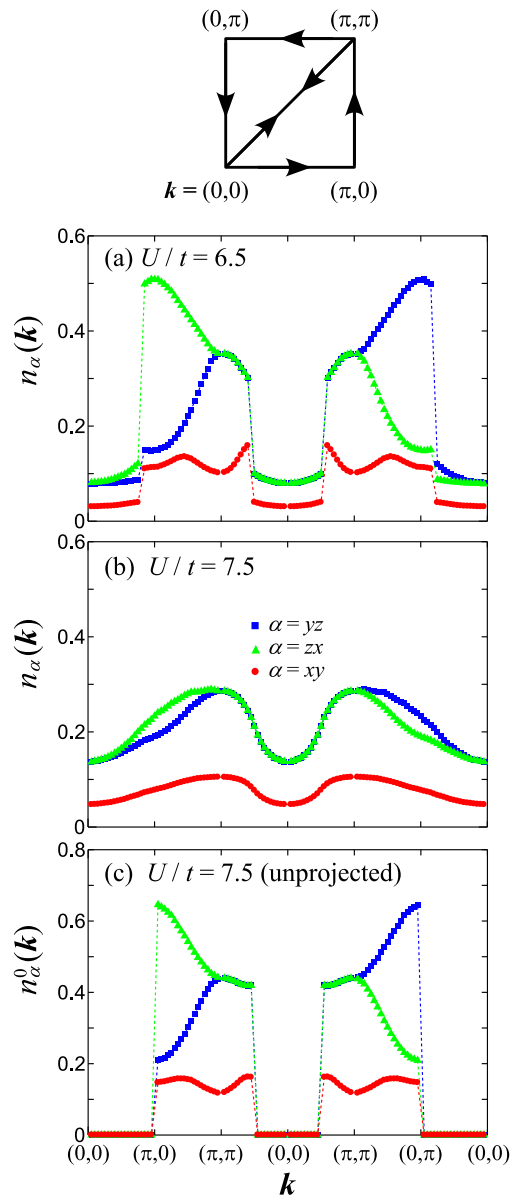}
\caption{\label{fig3} 
(color online) 
Hole momentum distribution functions $n_{\alpha}(\bm{k})$ for (a) $U/t=6.5$ and (b) $U/t=7.5$ with ($\lambda /t, J/U)$=(1.028, 0.0) 
and $L=20$.
For comparison, the unprojected momentum distribution function $n^0_{\alpha}(\bm{k})$ (see the text for definition) for $U/t=7.5$ is 
also shown in (c). The momentum pass taken in the first Brillouin zone is indicated in the upper panel. 
} 
\end{center}
\end{figure}

The metal-insulator transition is identified by the disappearance of discontinuities in the momentum distribution function.
For convenience, we calculate the momentum distribution function of holes, 
\begin{equation}
n_{\alpha}(\bm{k})=\frac{1}{2}\sum_{\sigma} \frac{{\langle\Psi|  c_{\bm{k}\alpha\sigma}c^{\dagger}_{\bm{k}\alpha\sigma} |\Psi\rangle}}
{{\langle\Psi|\Psi\rangle}},
\end{equation}
where $\alpha$ denotes the three $t_{2g}$ orbitals $d_{yz}$, $d_{zx}$, and $d_{xy}$.~\cite{note2}
Figure~\ref{fig3} shows $n_{\alpha}(\bm{k})$ for $U/t=6.5$ and 7.5 with $(\lambda /t, J/U)=(1.028, 0.0)$. 
Although the Hamiltonian $H$ and the variational wave function $|\Psi\rangle$ have a fourfold rotational symmetry, 
the $d_{yz}$ and $d_{zx}$ components have a strong one-dimensional character and each of them does not have a fourfold 
rotational symmetry by itself. Therefore, as indicated in the upper panel of Fig.~\ref{fig3}, 
we take the momentum pass in the first Brillouin zone through 
$(0,0)\to(\pi,0)\to(\pi,\pi)\to(0,0)\to(\pi,\pi)\to(0,\pi)\to(0,0)$ to show the behavior of each component more clearly.
Figure~\ref{fig3} (a) clearly exhibits discontinuities in $n_{\alpha}(\bm{k})$ for $U/t=6.5$, indicating the presence of well defined 
FS, i.e., a typical metallic behavior. To the contrary, in Fig.~\ref{fig3} (b) we see rather continuous variations of $n_{\alpha}(\bm{k})$ 
for $U/t=7.5$, a typical insulating behavior. 
It is interesting to note that the one-body part $|\Phi\rangle$ without the Gutzwiller-Jastrow projection for $U/t=7.5$ is 
metallic and its momentum distribution function $n^0_{\alpha}(\bm{k})=1/2\sum_\sigma\langle\Phi| c_{\bm{k}\alpha\sigma}c^{\dagger}_{\bm{k}\alpha\sigma} |\Phi\rangle/\langle\Phi|\Phi\rangle$ exhibit clear discontinuities as shown in Fig.~\ref{fig3} (c). 
Namely, the Gutzwiller and Jastrow factors remove these discontinuities in $n_{\alpha}(\bm{k})$ and make the system insulating, 
i.e., a ``true" Mott insulator is induced by electron correlations without breaking 
translational symmetry by any magnetic order. 

To confirm the presence of the metal-insulator transition in a different perspective, we examine the low-lying charge excitations 
by studying the charge structure factor. Following seminal work by Feynman and Cohen,~\cite{Feynman} 
the different behavior is expected in the charge structure factor $N(\bm{q})$ for a metal and an insulating states 
because $N(\bm{q})$ in the limit of $\bm{q}\rightarrow 0$ is related to the charge 
excitation gap. Assuming a wave function $|\Psi(\bm{q})\rangle=n_{\bm{q}}|\Psi\rangle$ for an excited state, 
the upper bound of the low-lying collective charge excitation energy $\Delta_C(\bm{q})$ is evaluated by
\begin{equation}
\Delta_C(\bm{q})=\frac {\langle\Psi(\bm{q})| H | \Psi(\bm{q})\rangle} {\langle\Psi(\bm{q})| \Psi(\bm{q})\rangle}
                             -\frac {\langle\Psi| H | \Psi\rangle} {\langle\Psi| \Psi\rangle},
\end{equation}
where $n_{\bm{q}}=\sum_{\bm{k}\alpha\sigma}c^{\dg}_{\bm{k}+\bm{q}\alpha\sigma}c_{\bm{k}\alpha\sigma}$ and $|\Psi\rangle$ is 
the optimized ground state wave function given in Eq.~(\ref{wf}). 
Because of the $f$-sum rule,~\cite{Feynman, Overhauser, Girvin} one can readily show that 
\begin{align}\label{delta_c(q)}
\Delta_C(\bm{q})=\frac{1}{2N(\bm{q})}&\sum_{\bm{k},\alpha,\sigma}\left[\varepsilon_{\alpha}({\bm{k}+\bm{q}})+\varepsilon_{\alpha}({\bm{k}-\bm{q}})-2\varepsilon_{\alpha}({\bm{k}}) \right] \notag \\
                                                               &\times \langle\Psi |c^{\dg}_{\bm{k}\alpha\sigma}c_{\bm{k}\alpha\sigma} |\Psi\rangle / \langle\Psi| \Psi\rangle,
\end{align}
where $N(\bm{q})$ is the charge structure factor calculated for the ground state wave function $|\Psi\rangle$, i.e., 
\begin{equation}
N(\bm{q})=\frac{1}{N}\frac{\langle\Psi| n_{-\bm{q}}n_{\bm{q}}|\Psi\rangle}{\langle\Psi| \Psi\rangle}.
\end{equation}
Taking the limit of $\bm{q}\rightarrow 0$ in Eq.~(\ref{delta_c(q)}), we find that $\Delta_C(\bm{q}) \propto |\bm{q}|^2/ N(\bm{q})$. 
Therefore, the system is metallic if $\lim_{\bm{q}\rightarrow 0}N(\bm{q}) \sim |\bm{q}|$ and 
insulating if $\lim_{\bm{q}\rightarrow 0}N(\bm{q}) \sim |\bm{q}|^2$.

\begin{figure}[thbp]
\begin{center}
\includegraphics[width=\hsize]{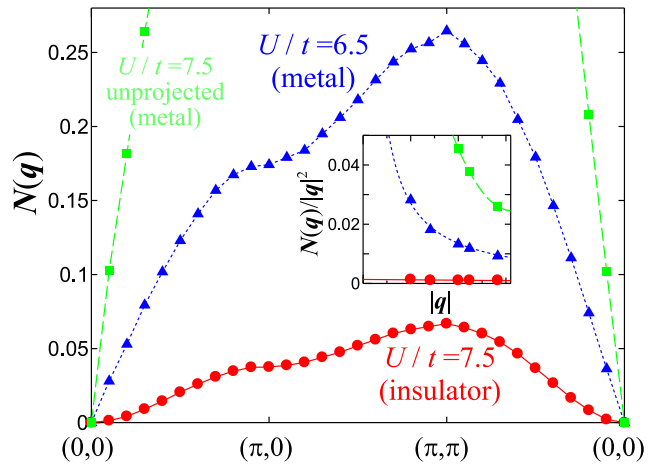}
\caption{\label{fig4} 
(color online) 
Charge structure factor $N(\bm{q})$ for $U/t=6.5$ and $7.5$ with ($\lambda /t, J/U)$=(1.028, 0.0) 
and $L=20$. 
For comparison, the unprojected charge structure factor 
$N_0(\bm{q})=\frac{1}{N}{\langle\Phi| n_{-\bm{q}}n_{\bm{q}}|\Phi\rangle}/{\langle\Phi| \Phi\rangle}$ for $U/t=7.5$ is 
also shown. 
(Inset) $N(\bm{q})/|\bm{q}|^2$ and $N_0(\bm{q})/|\bm{q}|^2$ for $|{\bm q}|\sim0$ are plotted. 
} 
\end{center}
\end{figure}

Figure~\ref{fig4} shows $N(\bm{q})$ for $U/t=6.5$ and 7.5 with $(\lambda /t, J/U)=(1.028, 0.0)$, the same parameter sets with 
the ones in Fig.~\ref{fig3}. Indeed, we can see, for ${\bm q}$ around $|\bm{q}|\sim 0$, $N(\bm{q})\sim|\bm{q}|$ behavior in 
the metallic state ($U/t=6.5$) and $N(\bm{q})\sim |\bm{q}|^2$ behavior in the insulating state ($U/t=7.5$). 
Namely, as shown in the inset of Fig.~\ref{fig4}, $\lim_{\bm{q}\rightarrow 0}N(\bm{q})/|\bm{q}|^2$ clearly exhibits 
the diverging (converging) behavior for the metallic (insulating) state, indicating the absence (presence) of a finite charge gap.
Consistently with the result shown in Fig.~\ref{fig3} (c), the one-body part $|\Phi\rangle$ alone without the Gutzwiller-Jastrow 
projection for $U/t=7.5$ exhibit the metallic behavior (Fig.~\ref{fig4}), indicating the importance of the Gutzwiller-Jastrow projection 
to describe an insulating state without breaking the translational symmetry.

\begin{figure}[t!]
\begin{center}
\includegraphics[width=0.9\hsize]{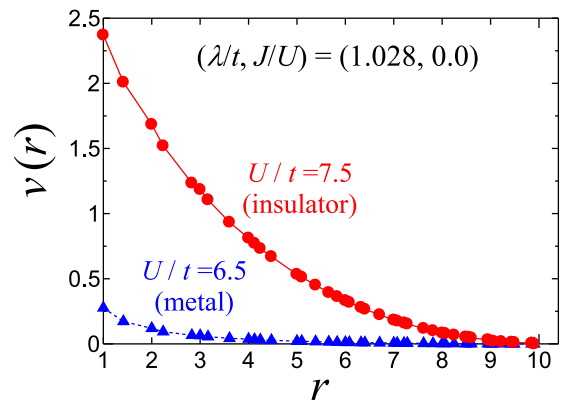}
\caption{\label{fig5} 
(color online) 
Distance $r$ dependence of charge Jastrow factor $v(r)$ in $P_{\rm J_c}$ [Eq.~(\ref{jastrow})] for $(\lambda/t,J/U)=(1.028,0.0)$ 
and $L=20$. The values of $U/t$ used are indicated in the figure.
} 
\end{center}
\end{figure}

Indeed, the insulating state is properly described by the paramagnetic wave function $|\Psi\rangle$ 
with a long-range charge Jastrow factor. 
Figure~\ref{fig5} shows the optimized charge Jastrow factor $v(r=|\bm{r}_i-\bm{r}_j|)$ in both metallic and insulating states
for ($\lambda /t, J/U)$=(1.028, 0.00). Since the wave function $|\Psi\rangle$ represents the same state even 
when an arbitrary constant is added to $v(r)$, 
we fix $v(r)=0$ for the longest distance $r(=L/2)$ considered here. As shown in Fig.~\ref{fig5}, 
not only the short-range terms but also the long-range terms give significant contribution in the insulating state, 
while $v(\bm{r})$ is much smaller and rapidly decays in the metallic state.
This indicates that the long-range charge correlation is essential for describing the Mott insulator, although the Coulomb interactions 
themselves in $H$ are only short-ranged. This is known in a single-orbital Hubbard model~\cite{Capello} and here we demonstrate 
that it is also the case in a multi-orbital Hubbard model with a large SOC.

\begin{figure}[t!]
\begin{center}
\includegraphics[width=0.65\hsize]{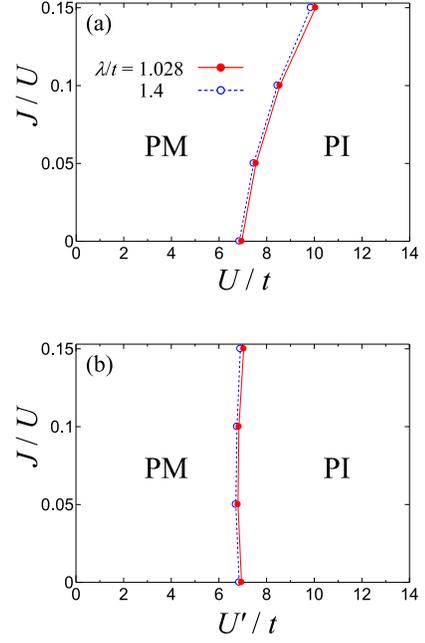}
\caption{\label{fig6} 
(color online) 
Ground state phase diagrams within a paramagnetic state in (a) $U/t$--$J/U$ and (b) $U'/t$--$J/U$ planes for 
$\lambda/t=1.028$ (solid lines) and $1.4$ (dashed lines). 
PM and PI denote a paramagnetic metal and a paramagnetic insulator, respectively. Note that $U=U'+2J$ is imposed. 
}
\end{center}
\end{figure}

Systematically studying $n_{\alpha}({\bm k})$ and $N({\bm q})$ for different values of $U/t$ and $J/t$, we obtain in Fig.~\ref{fig6} (a) 
the ground state phase diagram within the paramagnetic state for two different values of $\lambda/t=1.028$ and $1.4$. 
It is expected that larger $\lambda$ favors the insulating state 
because the energy split between the $J_{\mathrm{eff}}=1/2$ originated band and the 
$J_{\mathrm{eff}}=3/2$ originated bands becomes larger and thus the band overlap between the $J_{\mathrm{eff}}=1/2$ and 
$J_{\mathrm{eff}}=3/2$ bands becomes smaller (see Fig.~\ref{fig1} and Fig.~\ref{fig2}), which makes it easier to open the insulating gap with less electron correlations. 
Although this intuitive expectation is valid, the critical value $U_{\mathrm{p-MIT}}$ of $U$ at which the paramagnetic 
metal-insulator transition occurs 
is not much affected by $\lambda$ used here: only slightly $U_{\mathrm{p-MIT}}$ decreases by 
$0.1$--$0.2\,t$ with increasing $\lambda$ for a given $J/U$ as shown in Fig.~\ref{fig6} (a). 
Instead, we find that $U_{\mathrm{p-MIT}}$ is rather sensitive to $J/U$ in $U/t$--$J/U$ phase diagram: 
$U_{\mathrm{p-MIT}}$ increases with $J/U$, i.e., larger $J/U$ favoring the metallic state. 

To understand the $\lambda$ and $J/U$ dependence of $U_{\mathrm{p-MIT}}$, we should point out that the single-particle 
excitation gap in the atomic limit, i.e., $\varepsilon_{yz}(\bm{k})=\varepsilon_{zx}(\bm{k})=\varepsilon_{xy}(\bm{k})=0$, 
with $n=5$ is evaluated as
\begin{eqnarray}
\Delta_{\mathrm{c}}&=&E(d^4)+E(d^6)-2E(d^5) \label{delta_c} \\
&=&\frac{U+U'-J+2J'+6\lambda}{2} \nonumber\\
&&-\sqrt{\frac{(U-U'+J+2J'+2\lambda)^2}{4}+8\lambda^2} \\
&=&U-\frac{J}{2}+3\lambda-\sqrt{\frac{(2\lambda+5J)^2}{4}+8\lambda^2},  
\end{eqnarray}
where in the third line $J'=J$ and $U=U'+2J$ are assumed. $E(d^l)$ denotes the ground state energy for $l$ electrons. 
Thus, for a fixed $U$, $\Delta_{\mathrm{c}}$ monotonously increases with increasing $\lambda$ and monotonously decreases 
with increasing the Hund's coupling $J$. 
One can readily see that $\Delta_{\mathrm{c}}$ is not much sensitive to $\lambda$ because of the opposite 
contribution from the linear term and the square root term. Indeed, $\Delta_{\rm c}=U$ when $J=0$.~\cite{note3} 
On the other hand, $\Delta_{\mathrm{c}}$ is rather sensitive to $J$ because both the linear and the square root 
terms give the same (negative) contribution. In fact, $\Delta_{\rm c}=U-3J$ when $\lambda=0$.~\cite{note3} 
Namely, the Hund's coupling $J$ reduces the effective electron correlation and the metallic state is 
expected to be more favored with increasing $J$. These results should be relevant to understand the behavior of 
$U_{\mathrm{p-MIT}}$, at least, qualitatively,
which increases with increasing $J/U$ but is not much sensitive to $\lambda/t$. 
We should note that the effect of Hund's coupling to the metal-insulator transition has been discussed previously for 
multi-orbital Hubbard models without the SOC, 
where the charge gap in the atomic limit is indeed $U-3J$ at integer fillings with $n\ne m$ ($m$ being a number of orbitals) 
and is enhanced to $U+(m-1)J$ at half-filling, i.e., $n=m$.~\cite{deMedici1, deMedici2, Georges} 

It should be also noted that, as shown in Fig.~\ref{fig6} (b), the phase boundary between the metallic 
and insulating states is well scaled by $U'/t$ and insensitive to $J/U$ in $U'/t$--$J/U$ phase diagram. 
This seems contradict to the qualitative understanding given above as the charge gap in the atomic limit with $\lambda=0$ is 
$\Delta_{\rm c}=U'-J\,(=U-3J)$ rather than $U'\,(=U-2J)$. To better understand this, we consider the itinerant band effect. 
For this purpose, it is important to recall the electron configurations appearing in the right hand side of Eq.~(\ref{delta_c}). 
Using the notation defined in TABLE~\ref{Gutzwiller}, the charge gap $\Delta_{\rm c}$ is given as 
\begin{eqnarray}
\Delta_{\mathrm{c}}&=&E(d^4)+E(d^6)-2E(d^5) \nonumber \\
&=& E_{\rm I}(\left|\ua\da\;\;\ua\;\;\ua\right>) + E_{\rm I}(\left|\ua\da\;\;\ua\da\;\;\ua\da\right>) 
-2E_{\rm I}(\left|\ua\da\;\;\ua\da\;\;\ua\right>) \nonumber \\
&=&U'-J. \nonumber
\end{eqnarray}
Notice that because of the Hund's rule the 2-hole state for $d^4$ above is spin parallel and it has lower energy by $J'$ than 
a spin antiparallel 2-hole state 
$\left|\ua\da\;\;\ua\;\;\da\right>$, i.e., $E_{\rm I}(\left|\ua\da\;\;\ua\;\;\da\right>)-E_{\rm I}(\left|\ua\da\;\;\ua\;\;\ua\right>)=J'$ 
(see TABLE~\ref{Gutzwiller}). 
When the itineracy of electrons is taken into account, the antiferromagnetic correlation between the nearest neighbor sites 
is enhanced and the electron hopping between the nearest neighbor sites, which mainly contributes to the charge gap, 
competes with the Hund's rule. Therefore, considering the itinerant band effect, it would be natural to take the spin antiparallel 
2-hole state for $d^4$ and a rough estimation for the effective charge gap $\Delta_{\rm c}'$ is 
$\Delta_{\rm c}' \approx E_{\rm I}(\left|\ua\da\;\;\ua\;\;\da\right>) + E_{\rm I}(\left|\ua\da\;\;\ua\da\;\;\ua\da\right>) 
-2E_{\rm I}(\left|\ua\da\;\;\ua\da\;\;\ua\right>) =U'$.
Indeed, as shown later, the $H_{U'}$ term in $H_{\rm I}$ [Eq.~(\ref{int})] 
mainly contributes to the total energy and dominates the energy gain mechanism (see Fig.~\ref{fig12}). 
The details are discussed in the next subsection.

Next, we study the $U/t$ dependence of the generalized double occupancy $D$ defined by 
\begin{equation}
D = \frac{1}{N}\sum_{i}\frac{\langle\Psi|n_i^2|\Psi\rangle}{{\langle\Psi| \Psi\rangle}}.
\end{equation}
This is an extension of the double occupancy $D^{(1)}=(1/N)\sum_i\left<n_{i\ua}n_{i\da}\right>$ for a single-orbital system.
In a multi-orbital system, we should consider not only diagonal elements but also off-diagonal elements 
such as $n_{i\alpha\sigma}n_{i\beta\sigma'}(\alpha\neq\beta)$,
and these quantities depend on the choice of bases. 
Therefore, we take the sum with respect to spin and orbital indices, $n_i=\sum_{\alpha\sigma}n_{i\alpha\sigma}$, i.e., 
local density, for the generalized double occupancy $D$, which is basis invariant. 
For convenience, here, we calculate the generalized double occupancy of holes, 
$D^{\mathrm{hole}}={\frac{1}{N}}\sum_i\langle\Psi| (n^{\mathrm{hole}}_i)^2|\Psi\rangle/\langle\Psi|\Psi\rangle=D-24$, 
with $n^{\mathrm{hole}}_i=\sum_{\alpha}(2-n^{\alpha}_i)=6-n_i$. 
One can readily show that $D^{\mathrm{hole}}=1$ for $U\to\infty$ where the charge fluctuation is completely frozen. 
In the non-interacting limit with $U=J=0$, $D^{\mathrm{hole}}=1.549$ ($1.5$) for $\lambda=1.028t$ ($1.4t$), 
noting that the $J_{\rm eff}=3/2$ band with $m=2$ is partially (completely) occupied (see Fig.~\ref{fig1} and Fig.~\ref{fig2}). 

\begin{figure}[t!]
\begin{center}
\includegraphics[width=\hsize]{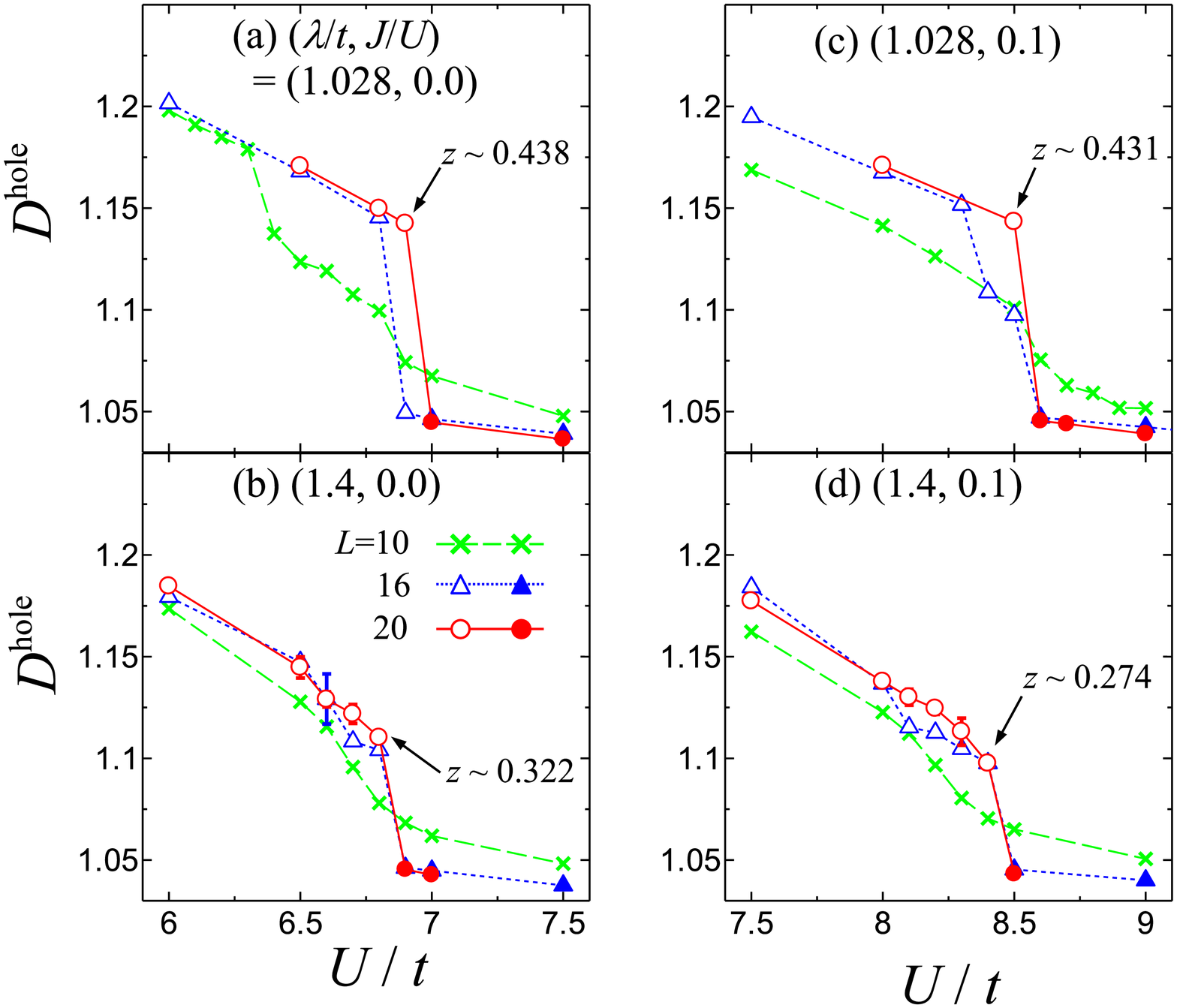}
\caption{\label{fig7} 
(color online) 
$U/t$ dependence of the generalized double occupancy of hole $D^{\mathrm{hole}}$ 
for different sets of parameters $(\lambda/t,J/U)$ indicated in 
the figures. The results for $L=16$ and 20 are denoted by triangles and circles, respectively, where open (solid) symbols 
represent the metallic (insulating) state. The results for $L=10$ are also plotted by crosses, from which 
the transition point is difficult to determine. The renormalization factor $z$, estimated from the jump of the momentum 
distribution function $n({\bm k})=\sum_\alpha n_\alpha({\bm k})$ for $L=20$, 
is indicated for the largest $U$ in the metallic phase.
}
\end{center}
\end{figure}

The results are summarized in Fig.~\ref{fig7} for different sets of $(\lambda/t,J/U)$. 
As shown in Fig.~\ref{fig7} (a), $D^{\mathrm{hole}}$ for ($\lambda /t, J/U)$=(1.028, 0.0) with 
$L=16$ and 20 exhibits clear discontinuity across 
the metal-insulator transition point, indicating that it is a first-order phase transition. 
For $L=10$, however, $D^{\mathrm{hole}}$ changes almost 
continuously and the transition is smeared out. The different behavior in $D^{\mathrm{hole}}$ is simply due to a 
finite size effect, i.e., the energy discretization near the Fermi level for $L=10$ exceeds the insulating gap and thus the 
$10\times10$ cluster fails to capture the transition correctly. This indicates that care should be taken for the size dependence of this quantity.
Indeed, as shown in Fig.~\ref{fig7} (c), even $L=16$ is not large enough to describe the transition in $D^{\mathrm{hole}}$ for 
($\lambda /t, J/U)$=(1.028, 0.1), although the first-order character of the transition seems to be partly captured for this $L$. 

\begin{figure}[t!]
\begin{center}
\includegraphics[width=0.85\hsize]{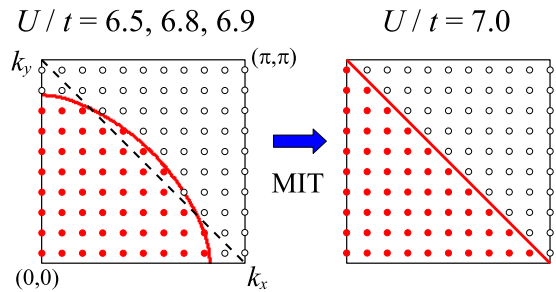}
\caption{\label{fig8} 
(color online) 
Evolution of the FS (red solid lines) in the one-body part $|\Phi\rangle$ of the paramagnetic wave functions for 
$(\lambda/t,J/U)=(1.028,0.0)$ and $L=20$. The momentum $\bm{k}=(k_x,k_y)$ inside (outside) the FS is denoted by red 
solid (black open) points. The Coulomb interaction $U/t$ is indicated in the figures. ``MIT'' denotes the metal-insulator transition. 
Only the first quadrant of the Brillouin zone is shown. 
} 
\end{center}
\end{figure}

\begin{figure}[t!]
\begin{center}
\includegraphics[width=0.8\hsize]{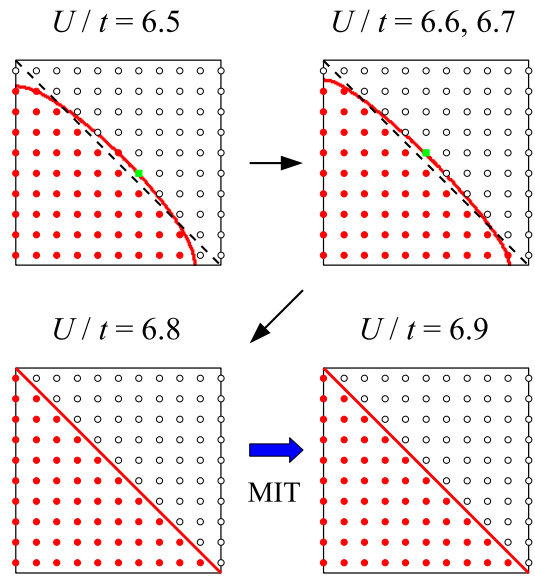}
\caption{\label{fig9}
(color online) 
Same as Fig.~\ref{fig8} but for $(\lambda/t,J/U)=(1.4,0.0)$. Green solid squares (and the symmetrically equivalent points) 
represent the ${\bm k}$ points which are only half occupied due to the degeneracy (i.e., open shell).
} 
\end{center}
\end{figure}

The $U/t$ dependence of $D^{\mathrm{hole}}$ for $\lambda/t=1.4$ is also shown in Figs.~\ref{fig7} (b) and \ref{fig7}~(d). 
Although the transition point of $U/t$ is almost the same with that for $\lambda/t=1.028$, the discontinuity at the transition point 
is smaller. This suggests that the first-order character of the transition is weakened for larger $\lambda/t$. 
This result can be understood by considering the shape of the FS in $|\Phi\rangle$ which is deformed from the original 
``bare'' FS in $H_0$ due to the electron correlations. 
When the Coulomb interactions increase, the FS in $|\Phi\rangle$ is spontaneously deformed to 
reduce the energy cost of Coulomb interactions 
at the expense of the kinetic energy. It turns out that in our three-orbital Hubbard model the FS in $|\Phi\rangle$ 
is deformed from a circular-like 
FS, as shown in Fig.~\ref{fig1} and Fig.~\ref{fig2}, to a tilted square FS as the Coulomb interactions increase, and 
in the Mott insulating state the FS in $|\Phi\rangle$ becomes almost perfectly tilted squared (see Fig.~\ref{fig8} and Fig.~\ref{fig9}). 
This is very similar to the case of the two-dimensional single-orbital Hubbard model with including up to the second nearest neighbor 
hoppings at half filling ($n=1$) where 
the FS of the corresponding $|\Phi\rangle$ is deformed to the tilted square shape satisfying the perfect nesting 
condition.~\cite{Tocchio} 
When the non-interacting FS of $H_0$ is far from the square (as in Fig.~\ref{fig1} with additional hole pockets), 
the metal-insulator transition occurs before the FS deformation is completed with increasing Coulomb interactions. 
Indeed, for ($\lambda/t, J/U$)=(1.028, 0.0), the $\bm{k}$-point occupation of the one-body part $|\Phi\rangle$ 
changes discontinuously through the metal-insulator transition (see Fig.~\ref{fig8}), thus causing the clear discontinuity 
in the one-body part $|\Phi\rangle$. 
On the other hand, for ($\lambda/t, J/U$)=(1.4, 0.0), the FS in $|\Phi\rangle$ is gradually deformed, as shown in Fig.~\ref{fig9}, 
with increasing Coulomb interactions 
and the FS deformation is completed at $U/t=6.8$ just before the metal-insulator transition occurs. 
In finite size calculations, it is difficult to determine the order of the transition when the FS deforms gradually. 
However, these results demonstrate that, despite almost the same $U_{\mathrm{p-MIT}}$, the behavior in the vicinity 
of the metal-insulator transition greatly depends on $\lambda$, 
which at the same time determines the shape of the original ``bare'' FS in $H_0$.

We also study the $J/U$ dependence of $D^{\mathrm{hole}}$ and find that the larger Hund's coupling $J/U$ makes 
the discontinuity smaller, although not drastically, as shown in Fig.~\ref{fig7}. 
This behavior can be understood by recalling that the larger $J/U$ increases $U_{\mathrm{p-MIT}}$ 
[Fig.~\ref{fig6} (a)] and thus a more renormalized metallic state becomes stable before entering the insulating phase. 
To quantify the degree of renormalization, the renormalization factor $z$ is estimated from the discontinuities in 
$n(\bm{k})=\sum_{\alpha}n_{\alpha}(\bm{k})$ at the Fermi momentum and the results are indicated in Fig.~\ref{fig7} for the largest $U$ 
in the metallic phase. 
Indeed, $z$ for the largest $U$ in the metallic phase becomes somewhat smaller with increasing $J/U$. 
On the other hand, $z$ greatly decreases with increasing $\lambda$, according with the stronger renormalization of the FS 
in $|\Phi\rangle$ towards the metal-insulator transition for larger $\lambda$ as discussed above. 
The smallest $z$ in Fig.~\ref{fig7} is $z\sim0.274$ for ($\lambda/t, U/t, J/U$)=(1.4, 8.4, 0.1),
which indicates that the effective electron mass ($m^*\approx m/z$) is about 4 times larger than the bare one ($m$).
This strongly renormalized metallic state has an interesting consequence, which is discussed in the next subsection.

\subsection{AF state and energy gain mechanisms}\label{af}

In the previous subsection, we have shown that there is the metal-insulator transition in the paramagnetic state. 
Here, we consider the AF orders in the wave function $|\Psi\rangle$ described in Sec.~\ref{vmc} 
and complete the ground state phase diagram by comparing the variational energies of paramagnetic and AF states. 
The results shown in this subsection are obtained for $L=20$, which is the maximal cluster size that we can treat 
with realistic computational time. 
Since we have confirmed that the in-plane AF state has always lower variational 
energy than the out-of-plane AF state for the parameter space considered here, 
the in-plane AF state is simply called ``AF state'' in the following. 

\begin{figure}[t!]
\begin{center}
\includegraphics[width=0.8\hsize]{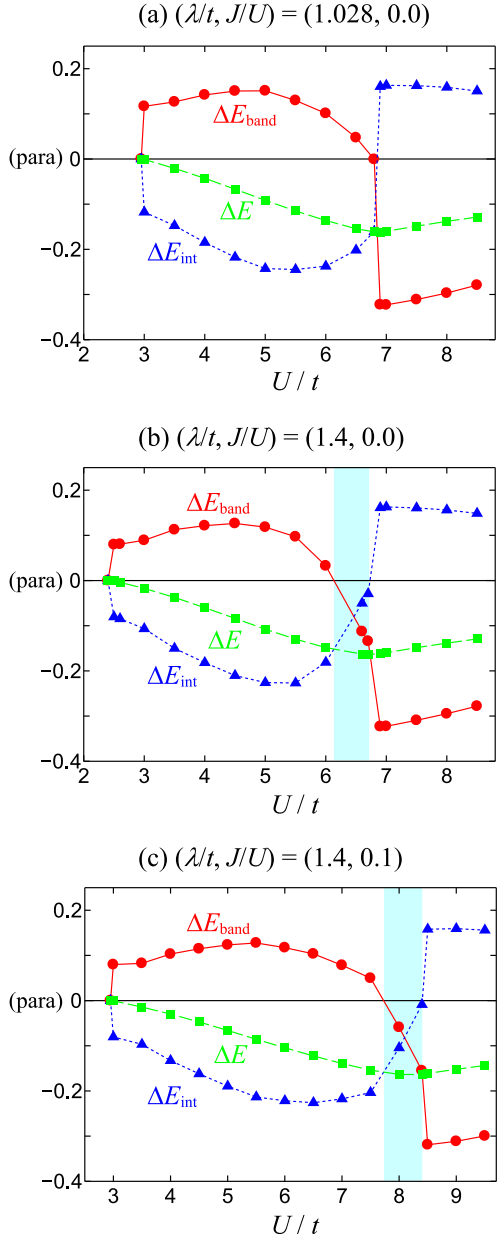}
\caption{\label{fig10} 
(color online) 
$U/t$ dependence of $\Delta E$, $\Delta E_{\mathrm{band}}$, and $\Delta E_{\mathrm{int}}$ for (a) ($\lambda /t, J/U)$=(1.028, 0.0), 
(b) ($\lambda /t, J/U)$=(1.4, 0.0), and (c) ($\lambda /t, J/U)$=(1.4, 0.1). Blue-shaded regions indicate the intermediate region where
$\Delta E_{\mathrm{band}}$ and $\Delta E_{\mathrm{int}}$ are both negative.
} 
\end{center}
\end{figure}

To complete the ground state phase diagram, first we calculate the energy difference $\Delta E$ between the paramagnetic 
state and the AF state, 
\begin{equation}
\Delta E=E_{\mathrm{AF}}-E_{\mathrm{para}}, 
\label{cond}
\end{equation}
with varying $U/t$ and $J/U$. Here, $E_{\mathrm{AF}}$ ($E_{\mathrm{para}}$) is the variational energy of 
the optimized AF (paramagnetic) state for a given model parameter. 
The critical $U_{\rm AF}$ for the AF order is determined where $\Delta E =0$ at $U=U_{\rm AF}$. 
Systematically calculating $\Delta E$, we find that, with increasing $U/t$, the AF-paramagnetic transition occurs way 
before the metal-insulator transition occurs in the paramagnetic state discussed in the previous subsection, 
i.e., $U_{\mathrm{AF}}<U_{\mathrm{p-MIT}}$ for given $J/U$ and $\lambda$. For example, 
as shown in Fig.~\ref{fig10}, $U_{\mathrm{AF}}/t\sim3.0$ for $(\lambda /t, J/U)$=(1.028, 0.0), 
$2.4$ for $(\lambda /t, J/U)$=(1.4, 0.0), and $3.0$ for $(\lambda /t, J/U)$=(1.4, 0.1). 
These values are indeed much smaller than $U_{\mathrm{p-MIT}}/t$ for the same sets of $(\lambda /t, J/U)$ [see Fig.~\ref{fig6} (a)]. 

Furthermore, we find that a metal-insulator transition occurs simultaneously when the system becomes antiferromagnetically 
ordered: an AF metallic state has always higher variational energy than the AF insulating state. 
We also calculate the variational energies of superconducting states with different pairing symmetries~\cite{Watanabe2} 
and find that the one 
with $d_{x^2-y^2}$ symmetry has the lowest variational energy. However, its variational energy is always higher than 
the AF insulating state for $U>U_{\rm AF}$ or the paramagnetic metallic state for $U<U_{\rm AF}$. 
Therefore, the superconducting state is not the ground state for this electron density at $n=5$. 
The possibility of superconductivity away from $n=5$ has been discussed in the previous reports.~\cite{Wang,Watanabe2,Yang} 

Next, let us examine the stabilization mechanism of the AF insulating state 
$|\Psi_{\rm AF}\rangle$ over the paramagnetic state $|\Psi_{\mathrm{para}}\rangle$. For this purpose, we divide $\Delta E$ into two parts, 
\begin{equation}
\Delta E=\Delta E_{\mathrm{band}}+\Delta E_{\mathrm{int}}, \label{eg}
\end{equation}
where $\Delta E_{\mathrm{band}}$ is the contribution from the band energy, i.e., the kinetic and SOC terms, 
\begin{eqnarray}\label{dE_band}
\Delta E_{\mathrm{band}} 
&=& \frac {\langle\Psi_{\rm AF}|(H_{\rm kin}+H_{\rm SO})|\Psi_{\rm AF}\rangle} {\langle\Psi_{\rm AF}|\Psi_{\rm AF}\rangle} \nonumber\\
&-&\frac {\langle\Psi_{\mathrm{para}}|(H_{\rm kin}+H_{\rm SO})|\Psi_{\mathrm{para}}\rangle} {\langle\Psi_{\mathrm{para}}|\Psi_{\mathrm{para}}\rangle},
\end{eqnarray}
and $\Delta E_{\mathrm{int}} $ is the contribution from the Coulomb interaction energy, 
\begin{equation}
\Delta E_{\mathrm{int}} 
= \frac {\langle\Psi_{\rm AF}|H_{\rm I}|\Psi_{\rm AF}\rangle} {\langle\Psi_{\rm AF}|\Psi_{\rm AF}\rangle} 
-\frac {\langle\Psi_{\mathrm{para}}|H_{\rm I}|\Psi_{\mathrm{para}}\rangle} {\langle\Psi_{\mathrm{para}}|\Psi_{\mathrm{para}}\rangle}.
\label{dE_int}
\end{equation}
The $U/t$ dependence of $\Delta E$, $\Delta E_{\mathrm{band}}$, and $\Delta E_{\mathrm{int}}$ are summarized in Fig.~\ref{fig10}. 
For small $U/t$, the energy gain of the AF insulating state is due to the interaction energy, i.e., $\Delta E_{\mathrm{int}}<0$ 
but $\Delta E_{\mathrm{band}}>0$, indicating that this AF insulator is interaction-energy driven. Instead, for large $U/t$, 
the AF insulating state is stabilized by gaining the band energy, i.e., $\Delta E_{\mathrm{band}}<0$ but 
$\Delta E_{\mathrm{int}}>0$, indicating that this AF insulator is band-energy driven. 

\begin{figure}[t!]
\begin{center}
\includegraphics[width=0.9\hsize]{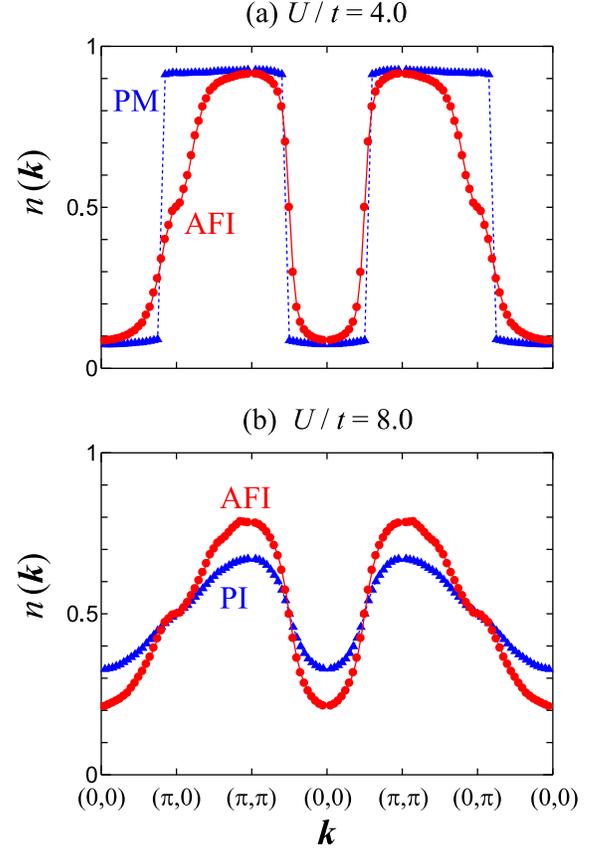}
\caption{\label{fig11} 
(color online) 
Total hole momentum distribution function $n({\bm k})$ for the paramagnetic and the AF states with $(\lambda/t,J/U)=(1.4,0.1)$.
As typical examples of the interaction-energy driven and the band-energy driven AF insulators, we choose (a) $U/t=4$ and 
(b) $U/t=8$.
PM, PI, and AFI denote paramagnetic metal, paramagnetic insulator, and AF insulator, respectively.
Momentum pass in the horizontal axis is the same as in Fig.~\ref{fig3}. 
Notice that, unlike $n_\alpha({\bm k})$ shown in Fig.~\ref{fig3}, the total hole momentum distribution is fourfold rotational 
symmetric. 
} 
\end{center}
\end{figure}

The change of the energy gain mechanism is also inferred in the momentum distribution function. 
Figure~\ref{fig11} shows the total hole momentum distribution function 
$n({\bm k})=\sum_\alpha n_\alpha({\bm k})$ for two different values of $U/t$ as typical examples for the interaction-energy driven and 
band-energy driven cases. Comparing $n({\bm k})$ for the paramagnetic and the AF states in Fig.~\ref{fig11}, it is suggestive that 
the band energy is lost when the paramagnetic state becomes the AF state for small $U/t$ [Fig.~\ref{fig11} (a)], while it is gained 
once the paramagnetic state is turned to the AF state for large $U/t$ [Fig.~\ref{fig11} (b)]. 

It should be emphasized that the change of the energy gain mechanism is due to the change of the nature of the paramagnetic state: 
The paramagnetic state is metallic in the interaction-energy driven region [Fig.~\ref{fig11}(a)] and it is insulating 
in the band-energy driven region [Fig.~\ref{fig11}(b)]. 
Since the ground state AF insulating state is described by the same wave function, the evolution from the 
interaction-energy driven to the band-energy driven AF insulators with increasing $U/t$ should be considered as a crossover, 
not a phase transition, separating a weakly correlated and a strongly correlated regions. 
The similar crossover has been discussed in a single-orbital Hubbard model both for an AF state~\cite{Yokoyama1} 
and a superconducting state.~\cite{Maier,Yanase,Gull,Yokoyama2} 
Our results for the three-orbital Hubbard model are consistent with these previous reports.  

Furthermore, we find, between these two regions, an intermediate region where $\Delta E_{\mathrm{band}}$ and 
$\Delta E_{\mathrm{int}}$ are both negative for ($\lambda /t, J/U)$=(1.4, 0.0) and (1.4, 0.1), 
as indicated by blue shade in Figs.~\ref{fig10} (b) and \ref{fig10}~(c), 
although such a region is absent for ($\lambda /t, J/U$)=(1.028, 0.0) 
[Fig.~\ref{fig10} (a)]. 
In this intermediate region, the paramagnetic state is metallic and strongly renormalized with the almost perfectly tilted 
squared FS in $|\Phi\rangle$, as shown in Fig.~\ref{fig9}. 
Note also that the strongly renormalized paramagnetic 
metallic state does not appear for ($\lambda /t, J/U$)=(1.028, 0.0), where the first-order character of the metal-insulator transition 
is strong and the FS in the paramagnetic wave function is deformed abruptly through the transition (see Fig.~\ref{fig8}). 
We find semiquantitatively the same results even for the smaller system size using $L=16$, which suggests that the intermediate region 
as well as the resulting phase diagrams shown below in Fig.~\ref{fig13} 
and Fig.~\ref{fig14} are rather robust.

\begin{figure}[t!]
\begin{center}
\includegraphics[width=0.9\hsize]{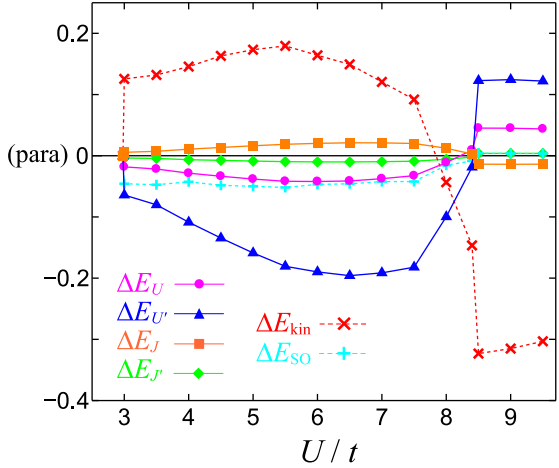}
\caption{\label{fig12} 
(color online) 
$U/t$ dependence of the energy difference between the paramagnetic state and the AF state for different terms in $H$. 
$\Delta E_{\mathrm{kin}}$, $\Delta E_{\mathrm{SO}}$, $\Delta E_{U}$, $\Delta E_{U'}$,
$\Delta E_{J}$, and $\Delta E_{J'}$ are indicated in Eqs.~(\ref{mh}) and (\ref{int}). 
The parameter used here is ($\lambda /t, J/U$)=(1.4, 0.1).  
} 
\end{center}
\end{figure}

\begin{figure}[t!]
\begin{center}
\includegraphics[width=0.9\hsize]{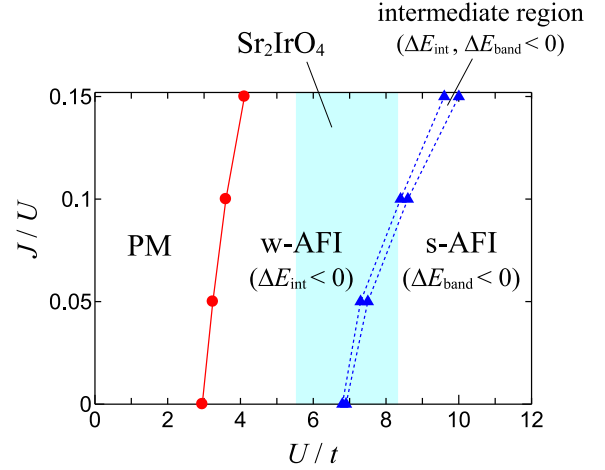}
\caption{\label{fig13} 
(color online) 
The ground state phase diagram including the AF state. PM, w-AFI, and s-AFI stand for paramagnetic metal, weakly correlated AF 
insulator, and strongly correlated AF insulator, respectively.
The intermediate region corresponds to a region where $\Delta E_{\rm band}$ and $\Delta E_{\rm int}$ are both negative. 
Blue shaded region indicates the values of $U/t$ relevant for Sr$_2$IrO$_4$. 
The SOC is set to be $\lambda /t=1.028$. 
} 
\end{center}
\end{figure}

\begin{figure}[t!]
\begin{center}
\includegraphics[width=0.9\hsize]{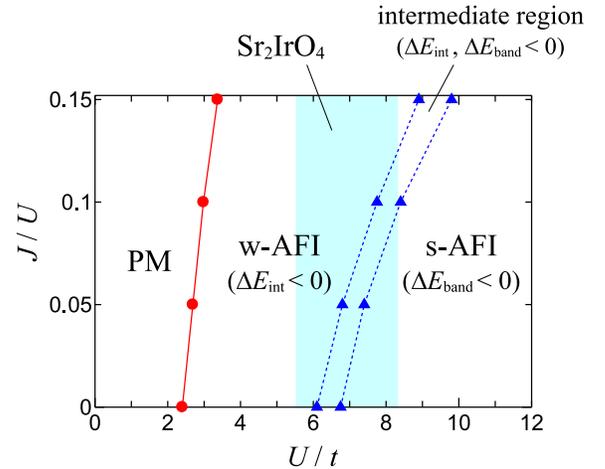}
\caption{\label{fig14} 
(color online) 
Same as Fig.~\ref{fig13} but $\lambda /t=1.4$. 
} 
\end{center}
\end{figure}

Finally, we examine the energy gain $\Delta E$ of the AF insulating state in more detail. 
Figure~\ref{fig12} shows $U/t$ dependence of each contribution in $\Delta E$ for ($\lambda /t, J/U)$=(1.4, 0.1). 
Here, $\Delta E_U$ and other quantities are defined similarly as in Eqs.~(\ref{dE_band}) and (\ref{dE_int}) for each term in $H$ 
[Eqs.~(\ref{mh}) and (\ref{int})]. 
We find that the inter-orbital Coulomb interaction $\Delta E_{U'}$ and the kinetic energy $\Delta E_{\mathrm{kin}}$ 
are the main contributions to $\Delta E_{\rm int}$ and $\Delta E_{\rm band}$, respectively, and thus determine the overall feature 
of the energy gain. The same tendency is also found for different values of $\lambda /t$ and $J/U$. 
This can explain the fact that the phase boundary of the paramagnetic metal-insulator transition is well scaled by $U'/t$ 
as shown in Fig.~\ref{fig6} (b).

To summarize this subsection, the ground state phase diagrams for $\lambda /t=1.028$ and $1.4$ are shown 
in Fig.~\ref{fig13} and Fig.~\ref{fig14}, respectively. 
With increasing $U/t$, the ground state changes from the paramagnetic metal to the weakly correlated AF insulator (w-AFI) 
followed by the crossover to the strongly correlated AF insulator (s-AFI). 
The w-AFI and s-AFI are distinguished by the energy gain mechanism and there exists the intermediate 
region where $\Delta E_{\rm band}$ and $\Delta E_{\rm int}$ are both negative and the 
paramagnetic metallic state is strongly renormalized. 

Let us now discuss where Sr$_2$IrO$_4$ is located in the phase diagram.
The value of $U$ for Sr$_2$IrO$_4$ is approximately estimated within a range of $U=2$--$3$ eV,~\cite{Arita,Martins,Kim4,Comin} 
thus corresponding to $U/t=5.6$--$8.3$ with $t\approx0.36$ eV. This region is indicated in Fig.~\ref{fig13} and Fig.~\ref{fig14} 
by blue shade. 
For both values of $\lambda$, it is located around the intermediate region. 
Note also that the value of Hund's coupling is estimated as large as $J/U=0.06-0.07$ by the constrained RPA method.~\cite{Arita}
Therefore, we consider Sr$_2$IrO$_4$ to be a ``moderately correlated'' AF insulator where the band effect and the correlation effect are 
both important. The similar conclusion has been reached also in the recent mean-field analysis of multi-orbital Hubbard models 
for Sr$_{n+1}$Ir$_n$O$_{3n+1}$ with $n=1$, $2$, and $\infty$.~\cite{Carter2}
This dual nature is also indicated in the dynamical mean field theory calculation,~\cite{Arita} where the continuous phase transition 
from the paramagnetic metal to the AF insulator is found with decreasing temperature, suggesting the Slater-type insulating 
mechanism.  It is also pointed out that the substantial cooperation of Mott-type correlation effects induce strongly 
renormalized ``bad metallic'' behavior in the paramagnetic metallic region above the N{\'e}el temperature.~\cite{Arita} 
Although the finite temperature calculation can not be directly compared with our ground state calculation, 
the bad paramagnetic metallic region found above the N{\'e}el temperature can be regarded as the strongly renormalized metallic 
state which appears in the intermediate region of the ground state phase diagrams shown in Figs.~\ref{fig13} and \ref{fig14}.

\section{discussion and summary}\label{summary}

We shall now address the question whether Sr$_2$IrO$_4$ is a Slater-type or a Mott-type insulator. 
Although there have been several experimental reports concerning this issue, these results are still controversial. 
The temperature dependence of the resistivity shows no significant changes at the N{\'e}el temperature $T_{\rm N}$,~\cite{Chikara} 
strongly indicating that Sr$_2$IrO$_4$ is a Mott-type insulator. 
The insulating gap of $\sim0.62$ eV estimated from the scanning tunneling microscopy/spectroscopy (STM/STS) is 
unusually large for a Slater-type insulator and claimed to be a Mott gap.~\cite{Nichols,Dai} 
On the other hand, the temperature dependence of the gap below $T_{\rm N}$ seems to be consistent with 
the Slater-type behavior,~\cite{Kini,Arita,Li,Suga} although the pseudogap-like behavior above $T_{\mathrm{N}}$ should be 
discussed in more detail. 
Moreover, the time-resolved photocarrier dynamics experiment suggests that Slater and Mott characteristics coexist 
in Sr$_2$IrO$_4$.~\cite{Hsieh} 
These incompatible results among different experimental observations rather represent the unique character 
of Sr$_2$IrO$_4$.
As discussed in Sec.~\ref{af}, our VMC results indicate that Sr$_2$IrO$_4$ is a ``moderately correlated'' AF insulator 
located between a Slater-type and a Mott-type insulators. It is thus expected that both characteristic behaviors can be observed 
in different experiments. Further theoretical as well as experimental studies are highly desirable to understand the seemingly 
incompatible experimental observations. 

In summary, we have studied the three-orbital Hubbard model with a large SOC 
by using the VMC method to discuss the insulating mechanism of Sr$_2$IrO$_4$. 
First, we have shown that there is the metal-insulator transition within the paramagnetic state, which can be described by the 
long-range charge Jastrow factor. We have found that the underlying FS in the wave function is spontaneously deformed to the 
perfectly tilted squared shape in the paramagnetic insulating state. 
We have also shown the presence of strongly renormalized metallic state in the vicinity of the metal-insulator transition, 
where the FS of the wave function is also tilted squared. 
Next, we have incorporated the AF orders in the wave function and found that $U_{\rm AF}$ is much smaller than $U_{\mathrm{p-MIT}}$ 
for a given set of $(\lambda/t,J/U)$ and that the AF insulating state is always favored over the AF metallic state. 
We have then examined the stabilization mechanism of the AF insulating state over the paramagnetic state. 
Systematic calculations of the energy gain for the AF insulating state have revealed that the ground state changes from 
the interaction-energy driven, i.e, weakly correlated Slater-type, AF insulator to the band-energy driven, 
i.e., strongly correlated Mott-type, 
AF insulator with increasing the Coulomb interactions. We have also shown that, between these two regions, there exists 
the intermediate region where the energy gain mechanism is both interaction- and band-energy driven and where 
the paramagnetic state is strongly renormalized metal with the almost perfectly tilted squared FS in the wave function. 
Based on our results, we assign Sr$_2$IrO$_4$ to be located in the intermediate region between the weakly correlated and the 
strongly correlated AF insulators and thus we expect that Slater-like and Mott-like behaviors can be both observed in Sr$_2$IrO$_4$.

\section*{Acknowledgments}
The authors thank J. Akimitsu, M. Isobe, H. Okabe, S. Fujiyama, R. Arita, S. Biermann, W. Ku, 
and H. Yokoyama for useful discussions.
The computation has been done using the RIKEN Cluster of Clusters (RICC) facility and the facilities of the Supercomputer Center,
Institute for Solid State Physics, University of Tokyo.
This work has been supported by Grant-in-Aid for Scientific Research from MEXT Japan (Grants No. 24740251, 
No. 24740269, and No. 25287096) and in part by RIKEN iTHES Project.


\begin{thebibliography}{99}
\bibitem{note1} 
During the stochastic optimization for the one-body parameters in $\tilde H_0$, the electron occupation in ${\bm k}$ points 
generally changes when $|\Phi\rangle$ is chosen as the ground state of $\tilde H_0$. The change of electron configurations
often causes the stochastic optimization unstable. 
To avoid this, we perform the stochastic optimization with keeping an electron occupation in ${\bm k}$ points. 
Considering several possible electron occupations separately, we determine the optimized variational state with the 
lowest variational energy. 
Even though the electron occupation in ${\bm k}$ points is fixed during the stochastic optimization, we have found that the optimized 
variational state is always the ground state of $\tilde H_0$ with the optimized variational parameters.
Note that this problem does not occur when the Gutzwiller-Jastrow parameters are stochastically optimized.
\bibitem{note2}
The results shown here are calculated for $L=20$. However, we have checked that the results are semiquantitatively the same 
for different system sizes of $L=14, 16$, and $18$.
\end{thebibliography}
\end{document}